\documentstyle[preprint,aps,epsf,epsfig,12pt]{revtex}
\tighten
\begin{document} 
\title{A Relativistic Chiral Quark Model For \\ Pseudoscalar Emission
From Heavy 
Mesons}
\author{J. L. Goity}
\address{
Department of Physics, Hampton University, Hampton, VA 23668, USA \\
and \\
Thomas Jefferson National Accelerator Facility  \\
12000 Jefferson Avenue, Newport News, VA 23606, USA.}
\author{W. Roberts}
\address{Department of Physics, Old Dominion University, Norfolk, VA
23529, USA 
\\
and \\
 Thomas Jefferson National Accelerator Facility \\
12000 Jefferson Avenue, Newport News, VA 23606, USA.}
\maketitle 
\begin{abstract}
The  amplitudes for one-pion mediated transitions between 
heavy meson excited states are obtained in the framework of the
relativistic
chiral quark model. The effective coupling constants to pions and the 
decay widths of excited 
  heavy mesons with $\ell\le 2$ for non-radially
excited,  
  and the $\ell=0$ radially excited mesons are presented for both
   charmed and beauty mesons. We also discuss the allowed decays of
strange
   excited heavy mesons by emission of a K-meson.

\vspace{5mm}

\flushright{JLAB-THY-98-36}
\end{abstract}
\thispagestyle{empty}
\pacs{  {\tt$\backslash$\string pacs\{\}} }
\setcounter{page}{1}

\section{Introduction}
It is known that the excited states of heavy mesons are of 
relevance in understanding the various weak decays of $B$ and $D$
mesons. One 
example
is the process $B\rightarrow D e \nu$ where the excited heavy
 meson states influence the slope of the Isgur-Wise form factor
$\cite{IW1}$ 
according to the
 Bjorken sum rule $\cite{Bj}$.$B_{\ell 4}$ decays are another example
where virtual
  excited states give rise to very important contributions to the total 
  rates as well as to the shape of the partial width in the $D\pi$
invariant mass
$\cite{goity1}$.
In most processes of interest these excited states  couple to the ground
state 
heavy meson   by emission of one or more pions. This is   the main
reason why a 
good understanding of these couplings is necessary and important. 

The properties of excited heavy mesons are also interesting in their own
right, as new insight into
strong interaction dynamics could be gained by studying such states. 
 Again, in this process  the knowledge of the pion couplings to the
excited mesons is crucial. Thus far, only the $D_1$ and $D_2$
states have been determined experimentally $\cite{Aleph1}$. 
These states  
  decay by emitting a pion in a D-wave, and for this reason they are
   relatively narrow and could be observed. Other excited states are
   likely to be substantially wider and thus harder to be observed. It
is important
   to provide estimates of these widths in order to judge the
feasibility
   of such observations. This is one of our objectives here.

In this work we study the couplings of pions to heavy mesons 
and their excitations in the framework of the relativistic version of
the
chiral quark model
of Georgi and Manohar $\cite{Ge-Ma}$. We also obtain the decay rates of
various
lower excitations of heavy mesons. Comparison with previous results
obtained in
non-relativistic models is made and, in particular, the important
relativistic
effects noticed in the S-wave decay widths are emphasized. The possible 
decays of excited heavy mesons with strangeness via K meson emission are
also 
discussed, and the interesting situation of some states that cannot
decay
in that way due to phase space is addressed.

Due to the heavy quark spin symmetry in the $M_Q\rightarrow \infty$
limit, 
 heavy meson states come in nearly
degenerate pairs of definite parity and with spins differing by one
unit.
In  such a pair the light degrees of freedom (light quark plus glue, or
in 
short,
the brown muck) are in a state of
angular momentum $j$ and parity $P$. Thus, all  pairs  have definite 
  $j^P$ quantum numbers. In theory, the implementation of the spin
symmetry 
is conveniently and elegantly    made by associating superfields 
$\cite{Falk1}$ with each pair. These superfields
are expressed in terms of Dirac matrices, and the tensor fields
(polarization 
tensors) required to 
describe the states in the doublet.
There  are two types of superfields depending on whether the parity is
 $P=(-1)^{j-\frac{1}{2}}$ or  $P=(-1)^{j+\frac{1}{2}}$. Defining 
 $k=j-\frac{1}{2}$, the superfields are
\begin{eqnarray}
P=(-1)^k &:&\nonumber\\
{\cal{H}}_-^{\mu_1...\mu_k}&=& \frac{1+\not{\! v}}{2\sqrt{2}}\;
\left\{ 
\vphantom{\frac{1}{2k+1}}Q_{j+1/2}^{\mu_1...\mu_{k+1}}\gamma_{\mu_{k+1}}\right.
-\sqrt{\frac{2k+1}{k+1}}\;\gamma_5\;Q_{j-1/2}^{\nu_1...\nu_{k}}\;
\left[\vphantom{\frac{1}{2k+1}}g_{\nu_1}^{\mu_1}...g_{\nu_k}^{\mu_k}\right.
\nonumber\\
&-&\frac{1}{2k+1}\left.\left.\vphantom{\frac{1}{2k+1}}\left(\gamma_{\nu_1}
(\gamma^{\mu_1}-v^{\mu_1})
g_{\nu_2}^{\mu_2}...g_{\nu_k}^{\mu_k}+...+
g_{\nu_1}^{\mu_1}...g_{\nu_{k-1}}^{\mu_{k-1}}
\gamma_{\nu_k}(\gamma^{\mu_k}-v^{\mu_k})\right)\right]\right\},\nonumber\\
&& \nonumber\\
P=(-1)^{k+1}&:&\nonumber\\
{\cal{H}}_+^{\mu_1...\mu_k}&=& \frac{1+\not{\! v}}{2\sqrt{2}}\;
\left\{ \vphantom{\frac{1}{2k+1}}
Q_{j+1/2}^{\mu_1...\mu_{k+1}}\gamma_5\,\gamma_{\mu_{k+1}}\right.
-\sqrt{\frac{2k+1}{k+1}}\; Q_{j-1/2}^{\nu_1...\nu_{k}}\;
\left[\vphantom{\frac{1}{2k+1}}
g_{\nu_1}^{\mu_1}...g_{\nu_k}^{\mu_k}
\right.\nonumber\\
&-&\frac{1}{2k+1}\left.\left.\vphantom{\frac{1}{2k+1}}
\left(\gamma_{\nu_1}(\gamma^{\mu_1}+v^{\mu_1})
g_{\nu_2}^{\mu_2}...g_{\nu_k}^{\mu_k}+...+
g_{\nu_1}^{\mu_1}...g_{\nu_{k-1}}^{\mu_{k-1}}
\gamma_{\nu_k}(\gamma^{\mu_k}+v^{\mu_k})
\right)
\right]
\right\}.
\end{eqnarray}
These superfields are traceless, totally symmetric tensors that  
are transverse with respect to the four-velocity $v_\mu$. They also
satisfy the constraint of being transverse to the Dirac $\gamma$
matrices. 
The spin-symmetry transformation law of the superfields is 
\begin{eqnarray}
{\cal{H}}^{\mu_1...\mu_k}&\rightarrow &
\exp\left(-i \vec{\epsilon}.\vec{S_v}\right) {\cal{H}}^{\mu_1...\mu_k},
\nonumber\\
S_v^j&=&i\,\epsilon^{jkl}\,[\not\!{e}_k,\, 
\not\!{e}_l]\;\frac{(1+\not\!{v})}{2},
 \end{eqnarray}
 where the $e_k^\mu$ are space like vectors orthogonal to the four
velocity.
 The conjugated superfields, which transform contravariantly under  
  spin-symmetry, are simply given  by ${\bar{\cal{H}}}^{\mu_1...\mu_k}=
  \gamma_0{\cal{H}}^{\dagger\;\mu_1...\mu_k}\gamma_0$. We are using here
  a normalization of the fields 
$Q^{\dagger}$ 
that is different from we used in our previous work $\cite{goity1}$. The current 
normalization corresponds to the standard non-relativistic
normalization.

The transition amplitudes of interest to us, namely the decay amplitudes
of an excited heavy hadron
into a heavy hadron and a light hadron, is proportional to 
\begin{equation}
(-1)^{(1/2+j^\prime+J_h+J)}\sqrt{(2J^\prime+1)(2j+1)}
\left\{\matrix{1/2&j^\prime&J^\prime\cr
J_h&J&j\cr}\right\},
\label{spincounting}
\end{equation}
which arises from spin counting arguments \cite{IW}.
The primed quantities above refer to the heavy daughter hadron, the
unprimed
ones refer to the parent, $J=j\pm 1/2$, $J^\prime=j^\prime\pm 1/2$,
$\{\}$ is a 
$6-J$ symbol and $J_h=S_h+\ell_h$. $S_h$ is the total angular
momentum of the light hadron $h$ and $\ell_h$ is the relative orbital
angular 
momentum between
$h$ and the heavy daughter produced in the decay. For the pion, $S_h=0$,
so
$J_h=\ell_\pi$. 

Consistent with this, one approach that has been used for the
description of
the emission of a soft pion is 
 HQET and ChPT combined into an effective theory $\cite{ChPTHM}$. In
this 
combined
effective
theory, the couplings of pions to ground state mesons, for instance, is
contained in the interaction term  
\begin{equation}
{\cal L}_{\chi}^{{\rm int}}=g \,{\rm Tr_{D}}\left( \bar{\cal 
H}_{-}\,\omega^{\mu}{\cal H}_{-}
\gamma_{\mu}\gamma_{5}\right),
\end{equation}
where
\begin{equation}\label{pseudovector}
\omega_{\mu}=\frac{i}{2}\,(u\partial_{\mu} u^{\dagger}-
u^{\dagger}\partial_{\mu} u)=\frac{1}{2F_{\pi}}\partial_{\mu}\Pi+\dots.
\end{equation}
In the above, ${\cal H}$ is an SU(3) triplet,  
  $\Pi=\pi^a \lambda^a$, the Gell-Mann matrices
are normalized to ${\rm Tr}( \lambda^a \lambda^b )=2 \delta^{ab}$,
$ u=\exp(i\Pi/2 F_\pi)$,
 $F_\pi=93$ MeV is the pion decay constant,  
and ${\rm Tr_{D}}$ denotes the trace over Dirac indices.
Similarly, it is 
straightforward to write down terms that involve  transitions  among
other spin-flavor multiplets.

In a more recent approach $\cite{Winston}$,
  the constraints of ChPT are
dispensed with in favor of allowing for the emission of pions which are
not 
necessarily soft.
 In the case of single pion emissions, for instance, this approach leads
to  
 an effective Lagrangian of the form
\begin{equation}
\frac{1}{2 F_\pi}\;{\rm Tr_{D}}\left[ 
\bar{{\cal{H}}^{\prime}}^{\mu_1...\mu_{k'}}   
\lambda^a   {\cal{H}}^{\nu_1...\nu_k}\; \gamma_5\;  A_{\mu_1...\mu_{k'}, 
\nu_1...\nu_k}
\pi^a \right],
\end{equation}
which is valid in the limit of SU(3). The factor of $1/(2F_\pi)$ is
included
here for dimensional reasons only. $A$ is a tensor of natural parity 
which is 
the
most general one that can be  built from Dirac matrices, the 
four velocity $v$ and the momentum of the pion. Implicit in this
interaction 
Lagrangian are the parity and angular momentum selection 
rules in the heavy quark limit, 
$J^P\rightarrow J'^{P'} +\ell_\pi$, 
$\ell_\pi+1=\mid P-P'\mid_{\rm mod\;2}$,
$\mid j-j'\mid \leq \ell_\pi\leq  j+j'$. 
For the transitions of interest to us we have
\begin{eqnarray}
(0^-,\; 1^-)\rightarrow (0^-,\; 1^-):&& ~~~~{\rm
P-wave~~~transition},\nonumber\\
A&=&\alpha_{ (0^-,\; 1^-)\rightarrow (0^-,\; 1^-)}\;
\gamma_\mu   \partial_\pi^\mu,\nonumber\\ \nonumber\\
g&=&-\alpha_{ (0^-,\; 1^-)\rightarrow (0^-,\; 1^-)},\nonumber\\
\nonumber\\
(0^+,\; 1^+)\rightarrow (0^-,\; 1^-):&& ~~~~{\rm
S-wave~~~transition},\nonumber\\
A&=&\alpha_{ (0^+,\; 1^+)\rightarrow (0^-,\; 1^-)}\;
\gamma_\mu   \partial_\pi^\mu,\nonumber\\ \nonumber\\
(1^+,\; 2^+)\rightarrow (0^-,\; 1^-):&& ~~~~{\rm
D-wave~~~transition},\nonumber\\
A^\nu &=& \frac{i}{\Lambda_\chi}\,\alpha_{ (1^+,\; 2^+)\rightarrow
(0^-,\; 
1^-)}\;
\gamma_\mu   \partial_\pi^\mu \partial_\pi^\nu,\nonumber\\ \nonumber\\
(1^-,\; 2^-)\rightarrow (0^-,\; 1^-):&& ~~~~{\rm
P-wave~~~transition},\nonumber\\
A^\nu &=& \alpha_{ (1^-,\; 2^-)\rightarrow (0^-,\; 1^-)}\;
 \partial_\pi^\nu,\nonumber\\ \nonumber\\
(1^-,\; 2^-)\rightarrow (1^+,\; 2^+):&& 
 ~~~~{\rm S-wave~~~transition},\nonumber\\
A_S^{\mu\,\nu}&=& \alpha_{ (1^-,\; 2^-)\rightarrow (1^+,\; 2^+)}^{(S)}\;
   g^{\mu \nu} v.\partial_\pi, \nonumber\\ 
   &&~~~~{\rm D-wave~~~transition},\nonumber\\
   A_D^{\mu\,\nu}&=& \frac{i}{\Lambda_\chi}\,\alpha_{ (1^-,\;
2^-)\rightarrow 
(1^+,\;
2^+)}^{(D)}\;
  {\cal{P}}_{\mu\nu\rho\sigma} \partial_\pi^\rho \partial_\pi^\sigma,
\label{couplings}
\end{eqnarray}
where ${\cal{P}}_{\mu\nu\rho\sigma}\equiv (g_{\mu\rho}-v_\mu v_\rho)
(g_{\nu\sigma}-v_\nu v_\sigma)-\frac{1}{3}  (g_{\mu\nu}-v_\mu v_\nu)
(g_{\rho\sigma}-v_\rho v_\sigma)$, and $\partial_{\pi}$ 
indicates the partial derivative has to act on the pion field. The 
dimensionless effective
couplings $\alpha$
will be extracted from our calculation in the chiral quark model. The
scale 
 $\Lambda_\chi$ will be taken to be $1$ GeV.
Both  approaches necessarily reproduce the results of Eqn. (3). Only
the 
effective
coupling constants are not known.   In this article, we
use the chiral quark model, along with relativistic wave functions for
the
heavy mesons, to arrive at estimates of these couplings. We will also
provide 
general results for the different
partial wave decay widths. In what follows we work in the rest frame of
the 
parent hadron. All expressions 
refer to that frame.



\section{The chiral quark model}

In the calculation of the decay amplitudes we use the chiral quark model 
as proposed by Georgi and Manohar$\cite{Ge-Ma}$. In this model 
the constituent light quarks acquire their masses from the spontaneous
chiral symmetry breaking plus a contribution from the current quark
masses
that explicitly break chiral symmetry. The latter can be neglected in
the 
case of the $u$ and $d$ quarks, but in the case of the $s$ quark it
ought to
be
included. This means that, while the axial vector coupling $g_A^q$   of
the
constituent quark can be taken to be the same for all three flavors, 
the effect of the SU(3) breaking by the quark masses must be included
in the Dirac equation of our quark model.  
Thus,  chiral symmetry   dictates the coupling of the Goldstone modes
(pions) to the constituent quarks. In terms of constituent quark fields,
the lowest order Lagrangian in the pion momentum expansion can be
expressed
as
\begin{eqnarray}
{\cal{L}}_{ChQM}&=&i\bar{q}\gamma^\mu\nabla_\mu\,q-
\bar{q}\tilde{m}_q \,q-\bar{q} V(x) q
+g_A^q\;
\bar{q}\gamma^\mu\gamma_5\omega_\mu\,q,\nonumber\\
\nabla_\mu &=&\partial^\mu+i\Gamma^\mu,~~~~~~\Gamma^\mu=\frac{i}{8
F_\pi}[\Pi,\,
\partial^\mu\Pi]+...,\nonumber\\
\omega^\mu &=&\frac{1}{2 F_\pi}\,\partial^\mu\Pi+\dots .\label{chquark}
\end{eqnarray}
Here $q$ represents the triplet of light quarks $(u,d,s)$, 
$\Gamma^\mu= \frac{i}{8 F_\pi} [\Pi,\;\partial^\mu \Pi]+...$,  and
$g_A^q$
 is the axial vector coupling of the constituent quark.
 The value of $g_A^q$  is not well known. Arguments in the large N$_c$
 limit show that $g_A^q=1+{\cal{O}}(1/{\rm N_c})$ $\cite{Weinberg}$,
 while an estimate of the nucleon axial coupling $g_A=1.255$ 
 obtained with the chiral quark model in a non-relativistic
approximation
 gives $g_A^q\simeq 0.75$ $\cite{Ge-Ma}$. Since relativistic effects
 must increase the ratio $g_A^q/g_A$, we will consider   $g_A^q$
 between the two mentioned values.
 We work in the limit where the heavy quark is regarded as 
infinitely heavy, and $V(x)$ represents the interaction of the light
valence 
quark with the heavy quark.  Here we use the  vector Coulomb plus linear
scalar potentials
with standard strengths for each piece.

We write the wave function of the light valence quark as
\begin{eqnarray}
\psi_{j\ell m} &=&  \left(
 \begin{array}{c} i  F(r)\; \Omega_{j\ell m}\\
 \\
  G(r) \;\Omega_{j\tilde{\ell} m}
 \end{array} \right), \nonumber\\
 \tilde{\ell} &=& 2j-\ell,
 \end{eqnarray}
 where the radial wave functions are real, and the spinor harmonics are
given by
 \begin{eqnarray}
\Omega_{j=\ell+\frac{1}{2}\,\ell\, m} &=& \left(
 \begin{array}{c} \sqrt{\frac{j+m}{2 j}}\;Y_{\ell\,m-\frac{1}{2}}\\
 \\
  \sqrt{\frac{j-m}{2 j}}\;Y_{\ell\,m+\frac{1}{2}}
 \end{array} \right), \nonumber\\
\Omega_{j=\ell-\frac{1}{2}\,\ell\, m} &=& \left(
 \begin{array}{c} \sqrt{\frac{j+1-m}{2
(j+1)}}\;Y_{\ell\,m-\frac{1}{2}}\\
 \\
  -\sqrt{\frac{j+m+1}{2 (j+1)}}\;Y_{\ell\,m+\frac{1}{2}}
 \end{array} \right).
\end{eqnarray}
Our conventions here correspond to those of Bjorken and Drell
$\cite{BjDrell}$.

The pion emission amplitudes require the calculation of the matrix
elements of
the time and spatial components of the axial-vector current operator to
which 
the 
pion
couples according to Eqn. (\ref{chquark}).
If $\vec{p}_{\pi}$ is the momentum of the pion,  the matrix elements of
the
time   component  of the axial current we need to consider are 
\begin{eqnarray}
 &~& \langle j' \ell' m'\mid 
  \gamma_0 \gamma_5 \exp(-i \vec{p}_{\pi}.\vec{r}) \mid j 
\ell  m \rangle
\nonumber\\
& = & 
\frac{4 \pi}{\sqrt{2 j'+1}}(-1)^{\ell-\ell'+j-j'}\sum_{\ell_\pi, m_\pi} 
(-i)^{\ell_\pi }\; Y^*_{\ell_\pi m_\pi}(\hat{p}_{\pi}) \; 
  \langle j m, \ell_\pi m_\pi\mid j'm'\rangle 
\nonumber\\
&\times& \int dr\, r^2\, \, j_{\ell_\pi}(p_{\pi}r) \;
\left\{
 F^\prime(r) G(r)\; \langle j' \ell'\mid\mid 
A_0^{\ell_\pi}
\mid\mid j\tilde{\ell}
\rangle\right.\nonumber\\
&-&
\left.
  G^\prime(r) F(r)\; 
\langle j' \tilde{\ell'}\mid\mid A_0^{\ell_\pi}
\mid\mid j\ell\rangle
\right\},
\end{eqnarray}
where
\begin{eqnarray}
\langle j'  \ell' \mid\mid A_0^{\ell_\pi}
\mid\mid j\ell\rangle
 &=&
 \frac{(-1)^{j+\ell_\pi+\frac{1}{2}}}{\sqrt{4 \pi}}\;
\left(
 \begin{array}{ccc} \ell_\pi & \ell &\ell'\\
 0&0&0    
 \end{array} \right)\;\nonumber\\
 &\times&\sqrt{(2 \ell+1)(2 \ell'+1)(2 \ell_\pi+1)(2 j+1)(2 j'+1)}
\left\{
 \begin{array}{ccc} \ell_\pi & \ell &\ell'\\
 \frac{1}{2}&j'&j   
 \end{array} \right\}.
\end{eqnarray}
Similarly, the matrix elements of the spatial components 
of the axial current are  
\begin{eqnarray}
& & \langle j' \ell' m' \mid \gamma_i \gamma_5 \exp(-i 
\vec{p}_{\pi}.\vec{r})\mid j \ell m \rangle
\nonumber \\ &=&
4\pi\;(-1)^{\ell-\ell'+j-j'}\sum_{\ell_\pi m_\pi}
(-i)^{\ell_\pi}\;Y^*_{\ell_\pi 
m_\pi}(\hat{p}_{\pi})\nonumber\\
&\times & \sum_{\ell^{*} m^{*}} \langle 1 i,\; \ell_\pi m_\pi\mid 
\ell^{*}m^{*}\rangle 
\langle j\, -\! m,\; j' m'\mid\ell^{*}m^{*}\rangle\nonumber\\
&\times& 
 \int dr\, r^2\,\;j_{\ell_\pi}(p_{\pi}r)
 \left\{
  F^\prime(r) F(r)\;
 \langle  j'  \ell' \mid\mid A^{\ell_\pi,\, \ell^*}
\mid\mid j\ell\rangle \right. 
\nonumber\\
&-&\left.   G^\prime(r) 
G(r)\;
   \langle  j'  \tilde{\ell'} \mid\mid A^{\ell_\pi ,\, \ell^*}
\mid\mid j\tilde{\ell}\rangle   
 \right\}  ,
\end{eqnarray}
where
\begin{eqnarray}
\langle j'  \ell' \mid\mid A^{\ell_\pi,\, \ell^*}
\mid\mid j\ell\rangle
 &=& \sqrt{\frac{3}{2 \pi}}\;
\sqrt{(2 \ell+1)(2 \ell'+1)(2 \ell_\pi+1)(2 j+1)(2 j'+1)}\nonumber\\
&\times& \left(
 \begin{array}{ccc} \ell & \ell_\pi &\ell'\\
 0&0&0    
 \end{array} \right) 
   \left\{
 \begin{array}{ccc} \frac{1}{2} & \ell &j\\
 \frac{1}{2}&\ell'&j' \\  
 1& \ell_\pi & \ell^*
 \end{array} \right\}.
\end{eqnarray} 
These expressions reflect the known fact that the time component of the
axial 
current
has matrix elements that are suppressed in the non-relativistic limit
($G(r), 
G^\prime(r)\to 0$), while the 
matrix elements
of the spatial components are not.

The general expression for the amplitudes is
\begin{eqnarray}
\langle J' J_3', j' \ell';\pi^a, \vec{p}_{\pi}\mid J J_3, j \ell\rangle
&=&\frac{i}{2 F_\pi} \xi^a g_A^q p_{\pi}^\mu \langle J' J_3', j'
\ell'\mid
A_\mu^a(\vec{p}_{\pi})\mid J J_3, j \ell\rangle,
\end{eqnarray}
where  $\xi^a$ is equal to $\sqrt{2}$ if $\pi^a=\pi^\pm$ and equal to 1
if $\pi^a=\pi^0$. 

After performing the coupling of the angular momentum $j$ of the light
quark to 
the 
heavy quark spin, one can express the above amplitude in a pion partial 
 wave decomposition
\begin{eqnarray}
\langle J' J_3', j' \ell';\pi^a, \vec{p}_{\pi}\mid J J_3, j \ell\rangle
&=&\frac{i}{2 F_\pi} \xi^a g_A^q\;\sum_{\ell_\pi m_\pi} (-i)^{\ell_\pi}
  \;Y^*_{\ell_\pi m_\pi}(\hat{p_{\pi}}) \; \langle \ell_\pi m_\pi, J
J_3\mid J' 
J'_3\rangle
 \nonumber\\
&\times& {\cal{A}}_{\ell_\pi  }(p_{\pi};J  ,j \ell; J'  , j' \ell').
\end{eqnarray}
Here the amplitudes ${\cal{A}}_{\ell_\pi  }(p_{\pi};J  ,j \ell; J'  , j'
\ell')$
are obtained from  (2.4) and (2.6) by performing the mentioned coupling
of  the 
heavy quark spin.
These amplitudes are, explicitly
\begin{eqnarray}
{\cal{A}}_{\ell_\pi  }&=&{\cal{A}}^0_{\ell_\pi  }
+{\cal{A}}^1_{\ell_\pi  },\nonumber\\
{\cal{A}}^0_{\ell_\pi }(p_{\pi};J  ,j \ell; J'  , j' \ell')&=&
4 \pi i (-1)^{\ell-\ell'+j-j'} E_{\pi}  (-1)^{J'+j'+\frac{1}{2}} \sqrt{2
J+1}
\left\{
 \begin{array}{ccc} \ell_\pi & j &j'\\
 \frac{1}{2}&J'&J   
 \end{array} \right\}\nonumber\\
 &\times & \int dr\, r^2\,\; j_{\ell_\pi}(p_{\pi}r)\left\{  F^\prime(r)
G(r) \langle 
j' \ell'\mid\mid 
A_0^{\ell_\pi}
\mid\mid j \tilde{\ell}\rangle\right.\nonumber\\
&-& \left.  G^\prime(r) 
F(r)
\langle j' \tilde{\ell'}\mid\mid A_0^{\ell_\pi}
\mid\mid j \ell\rangle\right\},
\nonumber\\
{\cal{A}}^1_{\ell_\pi  }(p_{\pi};J  ,j \ell; J'  , j' \ell')&=&
4 \pi(-1)^{\ell-\ell'+j-j'} p_{\pi}   \sqrt{2 J+1}\; \left\{ 
\begin{array}{ccc} \ell_\pi & j &j'\\
 \frac{1}{2}&J'&J   
 \end{array} 
 \right\}\;\nonumber\\ &\times&
 \sum_{\ell^*}
(- i)^{\ell^*-\ell_\pi}\; (-1)^{\ell_\pi+j-\frac{1}{2}+J'-\ell^*}
 \sqrt{(2 \ell^* +1) (2 \ell_\pi +1)}
\left(
 \begin{array}{ccc} 1 & \ell^* &\ell_\pi\\
 0&0&0  
 \end{array}\right)  \nonumber\\
&\times&   
 (-1)^{j+j'+\ell}
  \int dr\, r^2\, \; j_{\ell^*}(p_{\pi}r)
  \left\{
   F^\prime(r)F(r)\;
 \langle j' \ell'\mid\mid A^{\ell^*,\,\ell_\pi} \mid\mid j \ell \rangle
 \right.
 \nonumber\\
 &- &
 \left.  G^\prime(r)G(r)
 \langle j' \tilde{\ell'}\mid\mid A^{\ell^*,\,\ell_\pi} \mid\mid j
\tilde{\ell}
  \rangle \right\}.
\end{eqnarray}
These expressions are valid for emission of a light pseudoscalar from
any heavy
parent, in the chiral quark model. Note that both ${\cal
A}_{\ell_\pi}^0$ and 
${\cal A}_{\ell_\pi}^1$ explicitly
contain the factors expected from heavy quark spin symmetry and spin
counting
arguments. This in turn serves as a check that our results satisfy the 
constraints implied by heavy quark spin symmetry. Also note that 
in  this equation the reduced matrix element is evaluated with
$\ell^*$ and $\ell_\pi$ interchanged with respect to the expression
given in Eqn. (14).

With these expressions, the partial wave decay widths are obtained as
\begin{eqnarray}
\Gamma_{\ell_\pi}^a(J j \ell; J' j' \ell')=
\frac{2 J'+1}{2 J+1}\;\frac{{\xi^a}^2 {g_A^{q}}^2 \;p_{\pi}}{32 \pi^2
F_\pi^2}
\;\frac{M_{J'}}{M_J}\; \mid {\cal{A}}_{\ell_\pi  }(p_{\pi}; J, j \ell;
J', 
j'\ell')\mid^2.
\end{eqnarray}
Here the factor $\frac{M_{J'}}{M_J}$ arises from the normalization of
the heavy meson states.

Along similar lines one could consider the emission of a K meson 
in the decay of a strange excited heavy meson to a non-strange heavy
meson.
The flavor factor $\xi^a$ is in this case the same as for the emission
of 
a charged pion, i.e., equal to $\sqrt{2}$. Also one could look into 
$\eta$ meson emission.  In this case the factor  $\xi^a$ is equal to
$1/\sqrt{3}$ for non-strange heavy mesons and equal to $-2/\sqrt{3}$
for strange heavy mesons.

The results obtained above for ${\cal{A}}_{\ell_\pi  }$ can be checked
by performing the calculation
in a different way. Indeed, the term linear in $\partial_\mu\Pi$ in 
$g_A^q\,\bar{q}\gamma^\mu \gamma_5 \omega_\mu\,q$
in Eqn. (8) can be rewritten, after integration by parts and use of
the Dirac equation in the potential $V(x)$,
as
\begin{equation}
i \frac{g_A^q}{2 F_\pi}\, \bar{q} \left\{
\gamma_5,\;\tilde{m}_q+V(x)\right\}\,q.
\end{equation}
 When one uses the vector Coulomb plus scalar linear potential, 
 the result for ${\cal{A}}_{\ell_\pi  }$ becomes
 \begin{eqnarray}
 {\cal{A}}^1_{\ell_\pi  }(p_{\pi};J  ,j \ell; J'  , j' \ell')&=&
 4 \pi i\, (-1)^{(J'+1/2+\ell-\ell'+j)} \sqrt{2 J+1} 
 \; \left\{ 
\begin{array}{ccc} \ell_\pi & j &j'\\
 \frac{1}{2}&J'&J   
 \end{array} 
 \right\}\;\nonumber\\ &\times&
 \int dr \;r^2 j_{\ell_\pi}(p_\pi r) 2(\tilde{m}_q+K \, r)
 \left\{  F^\prime(r) G(r) \langle 
j' \ell'\mid\mid 
A_0^{\ell_\pi}
\mid\mid j \tilde{\ell}\rangle\right.\nonumber\\
&+& \left.  G^\prime(r) 
F(r)
\langle j' \tilde{\ell'}\mid\mid A_0^{\ell_\pi}
\mid\mid j \ell\rangle\right\}.
 \end{eqnarray}
 Clearly, since the two expressions obtained for ${\cal{A}}_{\ell_\pi 
}$ 
 only coincide when the Dirac equation for the bound light quark is
satisfied, 
 we can use them as a check that the numerical solutions for the wave
functions
 are sufficiently precise.

The effective couplings in Eqn. (\ref{couplings}) are also readily
obtained by
matching the amplitudes   in the  
superfield formalism with the amplitudes obtained in the chiral quark
model. For convenience
the matching is done with the latter multiplied   by $i^{\ell_\pi}$, and
we obtain: 
 \begin{eqnarray}
&&(0^-,\; 1^-)\rightarrow (0^-,\; 1^-):  ~~~~{\rm 
P-wave~~~transition}\nonumber\\
 \alpha_{ (0^-,\; 1^-)\rightarrow (0^-,\; 1^-)}&=&
 - g_A^q  \left\{ \int dr\, r^2\,
 \left[j_0( p_\pi r)\left(F^\prime(r)F(r)-\frac{1}{3} G^\prime(r) 
G(r)\right)-\frac{4}{3}
 j_2( p_\pi r) G^\prime(r) G(r) \right]  \right.\nonumber\\
 &+& \left. \frac{ E_\pi}{p_\pi} \int dr\, r^2\, j_1( p_\pi r)
\left(F^\prime(r) 
G(r)-
 G^\prime(r) F(r)\right)\right\};
 \nonumber\\ \nonumber\\
&&(0^+,\; 1^+)\rightarrow (0^-,\; 1^-):  ~~~~{\rm 
S-wave~~~transition}\nonumber\\
 \alpha_{ (0^+,\; 1^+)\rightarrow (0^-,\; 1^-)}&=& 
 - g_A^q \left\{  \int dr\, r^2\, j_0( p_\pi r)
  \left( F^\prime(r) G(r) - G^\prime(r) F(r) \right) 
\right.\nonumber\\
 &-&\left. \frac{p_\pi }{E_\pi}\int dr\, r^2\, j_1( p_\pi r) \left(
F^\prime(r) 
F(r)+ G^\prime(r) G(r) \right)  \right\};
 \nonumber\\ \nonumber\\
&&(1^+,\; 2^+)\rightarrow (0^-,\; 1^-): ~~~~{\rm
D-wave~~~transition}\nonumber\\
  \alpha_{ (1^+,\; 2^+)\rightarrow (0^-,\; 1^-)} &=& 
   \Lambda_\chi  \sqrt{3} g_A^q 
  \left\{ -\frac{E_\pi}{p_\pi^2} \int dr\, r^2\, j_2( p_\pi r)
\left(F^\prime(r) 
G(r)-G^\prime(r) F(r)\right)   \right.\nonumber\\
  &-& \left. \frac{1}{  p_\pi} \int r^2  dr \left[ j_1( p_\pi r)\left(  
F^\prime(r) 
F(r) -\frac{1}{5}G^\prime(r) G(r)\right)+
   \frac{6}{5} j_3( p_\pi r) G^\prime(r) G(r)\right]\right\};
 \nonumber\\ \nonumber\\
&&(1^-,\; 2^-)\rightarrow (0^-,\; 1^-):  ~~~~{\rm 
P-wave~~~transition}\nonumber\\
  \alpha_{ (1^-,\; 2^-)\rightarrow (0^-,\; 1^-)}&=&\;
   \sqrt{3} g_A^q \left\{ \frac{ E_\pi}{ p_\pi}  
  \int dr\, r^2\, j_1( p_\pi r) \left(F^\prime(r) G(r) 
-G^\prime(r) F(r)\right)  \right.\nonumber\\
&+&\left.   \int r^2  dr \left[ \frac{2}{3} j_0( p_\pi r) G^\prime(r)
G(r)-
  j_2( p_\pi r) \left(F^\prime(r) F(r)+\frac {1}{3} G^\prime(r) 
G(r)\right)\right]\right\};
 \nonumber\\ \nonumber\\
 &&(1^-,\; 2^-)\rightarrow (1^+,\; 2^+): \nonumber\\ 
&&~~~~{\rm S-wave~~~transition}\nonumber\\
  \alpha_{ (1^-,\; 2^-)\rightarrow (1^+,\; 2^+)}\mid_S &=& 
 - g_A^q \left\{  \int dr\, r^2\, j_0( p_\pi r) 
 \left(F^\prime(r) G(r)-  G^\prime(r) F(r) \right) \right. 
\nonumber\\
&-&\left. \frac{p_\pi }{E_\pi} \;\int dr\, r^2\, j_1( p_\pi r)
\left(F^\prime(r) 
F(r)+  
 G^\prime(r) G(r)\right)
  \right\};
  \nonumber\\ 
 &&~~~~{\rm D-wave~~~transition}\nonumber\\
 \alpha_{ (1^-,\; 2^-)\rightarrow (1^+,\; 2^+)}\mid_D&=& 
 -\Lambda_\chi\,   3 g_A^q \left\{- \frac{E_\pi}{  p_\pi^2} \int dr\,
r^2\, j_2( 
p_\pi r)
  \left(F^\prime(r) G(r) - G^\prime(r) F(r)\right)  
 \right.\nonumber\\
 &-& \left. \frac{2}{5  p_\pi} \int r^2  dr \left[ j_1( p_\pi r) 
 \left(F^\prime(r) F(r)+ G^\prime(r)
G(r)\right)\right.\right.\nonumber\\
 &+&
 \left. \left.\frac{3}{5 p_\pi } j_3( p_\pi r) \left(F^\prime(r) 
F(r)+ G^\prime(r)G(r)\right) \right]   \right\}. 
\label{consts}
\end{eqnarray}
 Note that in the non-relativistic limit ($G(r), G^\prime(r)\to 0$), the 
expression for 
 $\alpha_{ (0^-,\; 1^-)\rightarrow (0^-,\; 1^-)}$
 gives  the coupling of a pion to the ground state heavy mesons
$g=g_A^q$; 
  this is a factor two larger than the one  used in 
  $\cite{goity1}$, and is due to the different normalization of the 
  heavy meson fields we used there.  

In terms of the effective coupling constants, the decay widths 
for emission of a $\pi^0$ are
\begin{eqnarray}
\Gamma_{(0^-,1^-)\to(0^-,1^-)}&=& \frac{\mid
\alpha_{(0^-,1^-)\to(0^-,1^-)}\mid ^2}{24\pi F_\pi^2}
\frac{M_f}{M_i} \mid  \vec{p}_\pi \mid ^3, \nonumber\\
\Gamma_{(0^+,1^+)\to(0^-,1^-)}&=& \frac{\mid
\alpha_{(0^+,1^+)\to(0^-,1^-)}\mid ^2}{8\pi  F_\pi^2}
\frac{M_f}{M_i}E_\pi^2 \mid  \vec{p}_\pi \mid, \nonumber\\
\Gamma_{(1^+,2^+)\to(0^-,1^-)}&=& \frac{\mid
\alpha_{(1^+,2^+)\to(0^-,1^-)}\mid ^2}{24\pi  F_\pi^2 \Lambda_\chi^2}
\frac{M_f}{M_i}\mid  \vec{p}_\pi \mid^5, \nonumber\\
\Gamma_{(1^-,2^-)\to(0^-,1^-)}&=& \frac{\mid
\alpha_{(1^-,2^-)\to(0^-,1^-)}\mid ^2}{36 \pi F_\pi^2}
\frac{M_f}{M_i}\mid  \vec{p}_\pi \mid^3, \nonumber\\
\Gamma^{S-{\rm wave}}_{(1^-,2^-)\to(1^+,2^+)}&=& 
\frac{\mid \alpha_{(1^-,2^-)\to(1^+,2^+)}\mid ^2}{8\pi  F_\pi^2}
\frac{M_f}{M_i}E_\pi^2\mid  \vec{p}_\pi \mid, \nonumber\\
\Gamma^{D-{\rm wave}}_{(1^-,2^-)\to(1^+,2^+)}&=& 
\frac{\mid \alpha_{(1^-,2^-)\to(1^+,2^+)}\mid ^2}{144 \pi F_\pi^2
\Lambda_\chi^2}
\frac{M_f}{M_i}\mid  \vec{p}_\pi \mid^5, \nonumber\\
\end{eqnarray}
where $M_f$ is the mass of the heavy daughter hadron, and $M_i$ is the
mass of
the decaying parent. Except for the transitions within the ground 
state multiplet for which isospin breaking plays a crucial role, 
for all other cases
the total decay width into pions (ie, including neutral and charge pion
modes) is three times the width given in Eqn. (22).

\section{Results and Discussion}

The wave functions that we use for this calculation are obtained from
previous
work \cite{vanorden}. In \cite{vanorden}, the heavy meson spectrum is
obtained
by solving the one-body Dirac equation that arises from the Gross
reduction of
the Bethe-Salpeter equation. The potential used in that calculation
consisted
of a scalar confining and vector Coulomb potential, and the spectra
obtained
were in good agreement with experimental measurements. In addition, the
wave
functions and the Bethe-Salpeter formalism have been applied to a
calculation
of the Isgur-Wise function \cite{vanorden2}, and preliminary results
indicate
that the method compares well to others.

The wave functions in question have been obtained using two different
methods.
In the first, the two components of the Dirac equation are combined to
produce
a single second order differential equation, which is solved using
standard
numerical techniques. We refer to the wave functions obtained in this
way as
the `exact' wave function. In the second method, the wave functions are
expanded in a finite-sized basis of orthogonal functions, and a matrix 
diagonalization, as well as minimization in the energy eigenvalues are 
performed. The size of the basis is varied
to examine the sensitivity of our model predictions to the size of the
expansion 
basis. We have 
included the predictions for the
heavy meson spectrum for the states relevant for this work, in table 
\ref{table0}.

Our results for the coupling constants obtained in this model are shown
in table 
\ref{table2}. The integrals in Eqn. (\ref{consts}) depend on the
momentum of the 
light daughter
pseudoscalar, and are therefore `form factors' rather than `coupling
constants'. 
The results in the
table are obtained using $q^2=m_\pi^2$ for all but the intra-multiplet 
transitions of the $(0^-,1^-)$
multiplet, since most of these transitions are kinematically forbidden. 
For these, we use $q^2=10^{-4}$ GeV $^2$. Note that we show signs on the 
coupling constants in this table, but we
remind the reader that the signs we obtain depend on the sign
conventions used
for the wave functions, and are therefore not physically meaningful.
What is
meaningful is the relative sign between the S- and D-wave coupling
constants in
the fourth and fifth rows of the table.

We note that for most of the decays we consider,
the coupling constants show little sensitivity to the mass of the heavy
quark, with the
largest effect being of the order of 10\% as we go from $c$-flavored to 
$b$-flavored
mesons. The exception here are the decays of the radially excited
$(0^-,1^-)$ states. In addition, again with the exception of the radial
excitations of the ground state, SU(3) breaking effects are quite small.
The
reader is reminded that there are three possible sources of SU(3)
breaking in
our model. First, the strange quark mass used in obtaining the wave
functions is
different from that of the up and down quarks. Second, the (SU(3)
broken)
physical masses of the hadrons are used in calculating the coupling
constants,
and third, in the case of the decay widths, the (SU(3) broken) physical
masses
are used in the calculation of phase space. In the case of the coupling
constants, only the first two sources come into play, but we see from
our
results that the net effect is very small. It is nevertheless
interesting to
note that the heavy quark symmetry appears to be a better symmetry.

Comparison of these numbers with those that we reported earlier clearly
show the
importance of relativistic effects. This is especially evident for the 
transitions within the $(0^-,1^-)$ multiplet.
In the non-relativistic limit, this coupling constant, denoted $g$, has
the 
value of 1, in units of $g_A^{q}$. 
In the present
calculation, modulo a sign that arises from the definition, this
coupling 
constant ranges from 0.74 to 0.79, in the
same units.
Relativistic effects also become evident when we examine the decay
widths obtained in the model.

Our results for the decay widths of excited mesons, obtained using the
chiral
quark model, are shown in table \ref{table1}. In this table, all numbers
are in
MeV/$(g_A^{q\;2})$. The numbers in the columns labeled $D$ and $B$ arise
from 
pion
emission, while those in the columns labeled $D_s$ and $B_s$ arise from
kaon
emission. Blank entries in the table
correspond to decays that are kinematically forbidden. For
observed
states, we use the measured masses in calculating phase space. For
states that
have not yet been observed, we use the model predictions for the masses.
Our
estimates for the theoretical errors that we show are obtained by
varying the
masses of the parent hadrons by $\pm$20 MeV, and calculating the decay
rates at
the new masses.

The results that we report are obtained using the approximate wave
functions described above. For comparison, we have used 5, 10 and 15
basis
functions in the wave function expansion, and found that the results for
the
widths do not vary significantly. In fact, we
find a
change of about 2 MeV/$(g_A^{q\;2})$ $(\simeq 2\%)$ in the largest
widths we 
report, those for 
$0^+\to 0^-$. All variations we observe are of the order of or less than
$ 3\%$.

For the intra-multiplet transition of the $(0^-,1^-)$ multiplet 
our results   are
encouraging,
but suggest that $g_A^q$ should be less than unity. This is in keeping
with other
articles that suggest a value for this constant as small as 0.7. Using
the
latest total width reported for the $D^{*+}$ ($\Gamma_{D^{*\pm}}<$ 0.131
MeV), 
we find
$|g_A^q|<$ 0.95. Using this upper limit, and the published ratio of the
strong
to electromagnetic widths of the $D^{*0}$, we estimate that
$\Gamma_{D^{*0}}<$
0.094 MeV. The current experimental limit is $\Gamma_{D^{*0}}<$ 2.1 MeV.

The decays of the $(1^+,2^+)$ doublet would appear to require a value
for 
$|g_A^q|$ near the lower limit of 0.7, in order for
our results to be consistent with the latest widths reported in the
particle listings ($\Gamma_{D_1^0}=18.9^{+4.6}_{-3.5}$ MeV,  
$\Gamma_{D_1^\pm}=28\pm 8$ MeV, $\Gamma_{D_2^0}=23\pm 5$ MeV, 
$\Gamma_{D_2^\pm}=25^{+8}_{-9}$ MeV), \cite {PDG}. We remind the reader
that, 
except
for the decay within the ground-state multiplet, we do not consider
SU(2)
breaking effects in our calculation. 

There are large uncertainties on the
measured widths, and the total widths of the $D_1$ suggest that
$|g_A^q|>$ 1. It
is known, however, that the decays of the $D_1$ are not well described
in a
heavy quark symmetry (HQS) framework, unless $1/m_Q$ corrections are
taken into
account. In particular, an S-wave contribution to the decay of this
state can 
be expected \cite{falk}.
On the other hand, the decays of the $D_2$ have been well described in
the HQS
framework, without the need for $1/m_Q$ corrections. For instance, we
obtain
\begin{equation}
\frac{\Gamma_{D_2\to D\pi}}{\Gamma_{D_2\to D^*\pi}}=1.93,
\end{equation}
while the measured values are 2.3$\pm$0.6 for the neutral $D_2$, and
$1.9\pm
1.1\pm 0.3$ for the charged. If we crudely average the measure widths of
the 
$D_2$ for mixed charged
states to 24 MeV, our model would require $|g_A^q|$ =0.74. This would
then lead
to $\Gamma_{D^{*\pm}}=$ 0.082 MeV and $\Gamma_{D^{*0}}=$ 0.059 MeV.

As expected from previous work, our S-wave decay widths ($0^+\to 0^-$,
$1^+\to 
1^-$, $1^-\to 1^+$ and $2^-\to 2^+$) are relatively large, but much
smaller
than the values obtained in non-relativistic calculations, indicating
that
relativistic effects are significant for these decays. In fact, for all
of our
results, the decay widths that we obtain are significantly smaller than
the
numbers we obtained using a non-relativistic version of the chiral quark
model
\cite{goity1}. For the decay $0^+\to 0^-$, the result we obtain in the
present
model is about one-eighth of the width reported in \cite{goity1}. The
reduction
of S-wave amplitudes by relativistic effects has also been seen in other
model
calculations \cite{GK}. For the
radially excited pseudoscalar-vector doublet, the decay widths we obtain
are
about fifteen percent of those presented in \cite{goity1}: these states
may be much narrower than we originally predicted. It will be
interesting to see
what effects these differences in widths will have on the spectrum
expected in
$B_{\ell 4}$ decays.

For the strange mesons, we find that phase space suppresses or forbids
many of
the possible decays. This suggests that many of the $D_s$ and $B_s$
excited 
mesons will be quite narrow, as their primary mode of decay will be
radiative,
as well as through isospin-violating, OZI-rule-violating pion emission.
This 
`double-suppression' means that these decay widths should be quite
small, 
comparable to (or even smaller than) the radiative decay widths of these
states.

The results of this work derive from the strict use of the
chiral quark model,
and the experimental determinations of the widths would provide a
valuable test of this model.
In particular, we have seen that S-wave decay widths are particularly
sensitive to the model
chosen.  In summary, it would appear that the decays of excited heavy
mesons 
provide the best testing ground for the chiral quark model.


\section*{Acknowledgement}
WR gratefully acknowledges the hospitality and support of Universit\'e
Joseph 
Fourier, Grenoble, France, 
and of Institut des Sciences Nucl\'eaires, Grenoble, France, where part
of 
this work was done, and both authors thank the Department of Physics of
Florida State University, Tallahassee, and Dr. Simon Capstick in particular, 
for their kind hospitality 
while this work was being completed.
This work was supported by the
National Science Foundation through grant \# HRD-9633750 (JLG) and 
\# PHY 9457892 (WR), and by the Department 
of Energy  through contracts
DE-AC05-84ER40150 (JLG and WR) and DE-FG05-94ER40832 (WR).

 \begin{table}
\begin{tabular}{lllll}
State &\multicolumn{4}{c}{Mass (GeV)} \\ \cline{2-5}
 & $D$ & $B$ & $D_s$ & $B_s$ \\ \hline 
$0^-$ & 1.864 (1.85) & 5.279 (5.28) & 1.969 (1.94) & 5.369 (5.37) \\
$1^-$ & 2.007 (2.02) & 5.325 (5.33) & 2.114 (2.13) & 5.416 (5.43) \\
\hline

$0^+$ & 2.27 & 5.65 & 2.38 & 5.75 \\
$1^+$ & 2.40 & 5.69 & 2.51 & 5.79 \\ \hline

$1^+$ & 2.427 (2.41) & 5.69 & 2.535 (2.52) & 5.79 \\
$2^+$ & 2.459 (2.46) & 5.71 & 2.573 (2.58) & 5.82 \\ \hline

$1^-$ & 2.71 & 5.97 & 2.82 & 6.07 \\
$2^-$ & 2.74 & 5.96 & 2.86 & 6.07 \\ \hline

$0^{-\prime}$ & 2.50 & 5.83 & 2.61 & 5.93 \\
$1^{-\prime}$ & 2.62 & 5.87 & 2.73 & 5.97 \\ \hline

\end{tabular}
\caption{Masses of states obtained in the relativistic model. For
observed
states, the model masses are shown in parentheses.
\label{table0}}
\end{table}

\begin{table}
\begin{tabular}{lrrrr}
Decay &\multicolumn{4}{c}{(in units of $g_A^q$)} \\ \cline{2-5}
 & $D$ & $B$ & $D_s$ & $B_s$ \\ \hline 
 
$(0^+,1^+)\to (0^-,1^-)$ & $0.48$ &  $0.48$ & $0.46$ & $0.46$ \\

$(1^+,2^+)\to (0^-,1^-)$ & $-0.86$ &$ -0.88$ & $-0.84$ & $-0.86$ \\ 

$(1^-,2^-)\to (0^-,1^-)$ & $0.138$ & $0.138$ & $0.128$ & $0.128$ \\

$[(1^-,2^-)\to (1^+,2^+)]_S$ & $-0.50$ & $-0.50$ & $-0.48$ & $-0.48$ \\

$[(1^-,2^-)\to (1^+,2^+)]_D$ & $-1.08$ & $-1.08$ & $-1.04$ & $-1.04$ \\

$(0^-,1^-)^\prime\to (0^-,1^-)$ & $0.112$ & $0.142$ & $0.100$ &
$0.128$ \\

$(0^-,1^-)\to (0^-,1^-)$ & $-0.74 $& $-0.76$ & $-0.78$ & $-0.78$ \\\hline
\end{tabular}
\caption{Coupling `constants' obtained in the chiral quark model. In
the columns labeled $B$ and $D$, pions are emitted in the decays, while
in the columns labeled $D_s$ and $B_s$, the emitted mesons are kaons.
\label{table2}}
\end{table}

\begin{table}
\begin{tabular}{lcrrrr}
Decay & $\ell_\pi$ &\multicolumn{4}{c}{width (MeV/${g_A^q}^2$)} \\
\cline{3-6}
 & & $D$ & $B$ & $D_s$ & $B_s$ \\ \hline 
$0^+\to 0^-$ & 0 &  $125\pm 10$ & $133\pm 13$ & - & - \\
$1^+\to 1^-$ & 0 & $121\pm 10$ & $129\pm 14$ & - & - \\ \hline

$1^+\to 1^-$ & 2 & $17.7\pm 4.1$ & $12.5\pm 3.9$ & - & - \\
$2^+\to 0^-$ & 2 & $28.6\pm 3.7$ & $11.4\pm 2.7$ & $1.9\pm 1.0$ & - \\
$2^+\to 1^-$ & 2 & $14.8\pm 3.0$ & $9.9\pm 2.8$ & - & - \\ \hline

$1^-\to 0^-$ & 1 & $14.0\pm 0.7$ & $13.4\pm 1.2$ & $15.0\pm 1.0$ &
$10.5\pm 1.8$ 
\\
$1^-\to 1^-$ & 1 & $4.5\pm 0.4$ & $5.3\pm 0.6$ & $3.9\pm 0.6$ & $3.7\pm
0.7$ \\
$2^-\to 1^-$ & 1 & $15.4\pm 1.3$ & $15.1\pm 1.8$ & $14.9\pm 1.7$ &
$10.9\pm 2.2$ 
\\ \hline

$1^-\to 1^+$ & 0 & $52.8\pm 6.8$ & $57.5\pm 7.8$ & - & - \\
$1^-\to 1^+$ & 2 & $0.57\pm 0.21$ & $0.66\pm 0.36$ & - & - \\
$1^-\to 2^+$ & 2 & $0.26\pm 0.16$ & $0.40\pm 0.26$ & - & - \\
$2^-\to 1^+$ & 2 & $0.63\pm 0.27$ & $0.31\pm 0.19$ & - & - \\
$2^-\to 2^+$ & 0 & $52.3\pm 6.7$ & $45.9\pm 7.4$ & - & - \\
$2^-\to 2^+$ & 2 & $0.77\pm 0.40$ & $0.43\pm 0.32$ & - & - \\ \hline

$0^{-\prime}\to 1^-$ & 1 &$26.1\pm 1.8$ & $34.3\pm 2.2$ &
$0.13^{+1.40}_{-0.13}$ 
& $1.7^{+2.8}_{-1.7}$ \\
$1^{-\prime}\to 0^-$ & 1 &$10.6\pm 0.3$ & $13.5\pm 0.3$ & $9.5\pm 0.4$
& 
$5.5\pm 1.3$ \\
$1^{-\prime}\to 1^-$ & 1 &$22.4\pm 0.5$ & $25.3\pm 1.1$ & $9.5\pm 1.8$
& 
$5.2\pm 2.4$ \\ \hline

$D^{*+}\to D^+\pi^0$ & 1 & 0.045 & - & - & - \\
$D^{*+}\to D^0\pi^+$ & 1 & 0.101 & - & - & - \\
$D^{*0}\to D^0\pi^0$ & 1 & 0.065 & - & - & - \\ \hline

\end{tabular}
\caption{Decay widths of heavy mesons obtained in the chiral quark
model. In
the columns labeled $B$ and $D$, pions are emitted in the decays, while
in the
columns labeled $D_s$ and $B_s$, the emitted mesons are kaons. All
widths
shown are in MeV, and require a factor of ${g_A^q}^2$. Blank entries in
the
table correspond to decays that are kinematically forbidden.
\label{table1}}
\end{table}

\end{document}